\RequirePackage{fix-cm}
\documentclass[smallextended]{svjour3} 
\smartqed 
\usepackage{graphicx} 
\usepackage{graphics}

\usepackage{amsmath} 
\usepackage{enumerate}
\usepackage{booktabs}
\usepackage{multirow}
\usepackage{threeparttable}
\usepackage[latin1]{inputenc}
\usepackage{array}

\usepackage{mathptmx} 

\usepackage{subfigure} 
\usepackage{amssymb}
\usepackage{algorithm}
\usepackage{algorithmicx}
\usepackage{algpseudocode}
\usepackage{subfigure}
\usepackage[breaklinks,colorlinks,linkcolor=blue,citecolor=blue,urlcolor=blue]{hyperref}
\usepackage[numbers,sort&compress]{natbib} 
\usepackage{latexsym}
\begin{document}

\title{Modeling and Simulation of Practical Quantum Secure Communication Network}


\author{Yaxing Wang         \and
             Qiong Li         \and
             Qi Han         \and
             Yumeng Wang
}


\institute{Qiong Li \at
              School of Computer Science and Technology, Harbin Institute of Technology, Harbin, China \\
              Tel.: +86-451-86402861-868\\
              Fax:  +86-451-86402861-861\\
              \email{qiongli@hit.edu.cn}
}

\date{Received: date / Accepted: date}

\maketitle

\begin{abstract}
As the Quantum Key Distribution (QKD) technology supporting the point-to-point application matures, the need to build the Quantum Secure Communication Network (QSCN) to guarantee the security of a large scale of nodes becomes urgent. Considering the project time and expense control, it is the first choice to build the QSCN based on an existing classical network. Suitable modeling and simulation are very important to construct a QSCN successfully and efficiently. In this paper, a practical QSCN model, which can reflect the network state well, is proposed. The model considers the volatile traffic demand of the classical network and the real key generation capability of the QKD devices, which can enhance the accuracy of simulation to a great extent. In addition, two unique QSCN performance indicators, ITS (information-theoretic secure) communication capability and ITS communication efficiency, are proposed in the model, which are necessary supplements for the evaluation of a  QSCN except for those traditional performance indicators of classical networks. Finally, the accuracy of the proposed QSCN model and the necessity of the proposed performance indicators are verified by plentiful simulations results.

\end{abstract}

\newpage

\section{Introduction}
\label{sec:Introduction}
Based on the quantum key distribution (QKD) technology with the intrinsic characteristics of point-to-point (P2P) \cite{bennett2014quantum}, building a quantum secure communication network (QSCN) \cite{chen2018cryptanalysis, xu2015novel, chen2017controlled} with multiple QKD devices is a prevailing solution to overcome the limits of node scale and the communication distance \cite{elliott2005current, elliott2007darpa, alleaume2009topological, dianati2007architecture, sasaki2011field, chen2017experimental}. Since the state-of-the-art QKD technology has been able to support many applications at distances of several hundred kilometers in the P2P mode, those achievements have made it possible to establish a QSCN upon an existing complex classical network. For instance, the key distribution rate of one QKD device could reach 10 Mbps \cite{yuan201810} and 1Mbps \cite{dixon2010continuous} at the distribution distance of 10 km and 100 km, respectively. Additionally, the key distribution distance of one QKD device could reach 404 km in optical fiber \cite{yin2016measurement} and even reach 1200 km in free space \cite{liao2017satellite}. As to the existing QSCNs, the node scale has expanded from 6 nodes \cite{peev2009secoqc, sasaki2011field} to 56 nodes \cite{razavi2018introduction}, and the communication distance has extended from 19.6 kilometers \cite{peev2009secoqc} to 2000 kilometers \cite{razavi2018introduction}. With the growing scale and complexity of QSCNs, the aforehand modeling and simulation become crucial for the functional verification, the deployment optimization of QKD devices, the project time and cost control, the network quality assurance, etc. \cite{watanabe2006security, dianati2008architecture, diamanti2016practical}.

Unlike the field of classical networks, the modeling and simulation in the field of QSCN have not drawn many attentions \cite{li2015perfect, li2016quantum, xu2015network, maurhart2013new, yang2017qkd, mehic2017implementation}. In order to reduce the cost of communication resources, some efforts in quantum network coding \cite{li2015perfect, li2016quantum, xu2015network} have been made. However, the analysis of network performance was neglected in these researches. A simple trust relaying based \cite{salvail2010security, scarani2009security} QSCN simulation model was built by Yang et al. in 2017 \cite{yang2017qkd} to calculate the key consumption cost. In this work, the key generation capability was set to be endless when the key consumption of a pair of partners was evaluated, which is not realistic in a practical QSCN. Taking the limited key generation capability of QKD device into account, a relatively complete QSCN model \cite{mehic2017implementation} was proposed and implemented by Mehic et al. based on the Network Simulator-version 3 (NS-3) \cite{henderson2008network} in 2017. In the \cite{mehic2017implementation}, the key generation and traffic generation have been simulated to estimate network performance. However, the assumptions about the key generation capability and traffic demand are unreasonable. In addition, the network performance was only measured by routing cost and packet delivery ratio (PDR), which cannot indicate the characteristics of QSCN.

In a practical QSCN, the most important performance depends on whether the QKD key generation capability can meet the traffic demand \cite{gelfond2012key}. However, the modeling of traffic demand, key generation capability and their relationship have not been studied thoroughly in previous studies on QSCN. The main defects in the existing studies include the following. Firstly, the end-to-end (E2E) traffic demand is modeled by a constant which cannot describe the volatility of the network. Secondly, the P2P key generation capability is modeled by the performance of the QKD post-processing system without taking into account the effects of QKD optical system performance. Thirdly, there are still lack of suitable performance indicators to measure the relationship between the E2E traffic demand and P2P key generation capability. Fourthly, only one pair of terminal partners communicate in the simulation, which fails to reflect the whole status of a practical multi-party network.

In this paper, a practical QSCN model is proposed, in which the volatility of the E2E traffic demand is modeled by the Poisson stochastic process, and the P2P key generation capability is modeled by the GLLP theory. In addition, two performance indicators: ITS communication capability and ITS communication efficiency are proposed to evaluate the special all-round performance of a QSCN. Finally, a QSCN simulation in the view of the whole network is designed to verify the accuracy of the proposed QSCN model and the necessity of the proposed performance indicators.

The rest of this paper is organized as follows: Sec. \ref{sec:definition} gives the definition and basic characteristics of QSCN. Sec. \ref{sec:design} describes the practical QSCN model by proposes traffic generation module, key generation module and two performance indicators. Sec. \ref{sec:simulation} designs a simulation to analyze the network performance in detail based on the QSCN model. Sec. \ref{sec:conclusion} concludes this study and outlines the future works.

\section{Definition of quantum secure communication network}
\label{sec:definition}
\begin{figure}[htbp]
	\centering
	\includegraphics[scale=0.7]{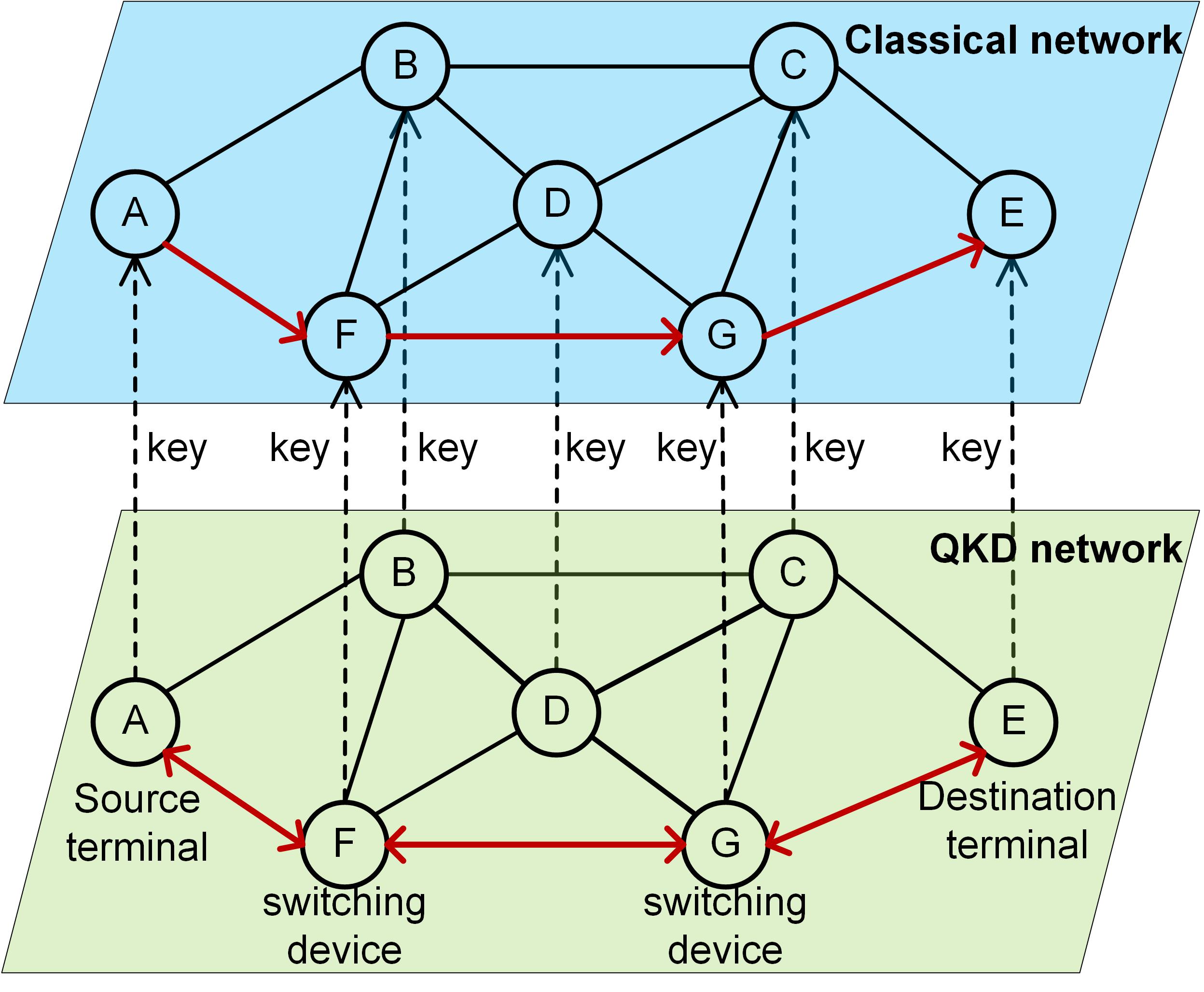}
	\caption{Hierarchical diagram of QSCN}
	\label{fig:network} 
\end{figure}

In many studies, both terms of QSCN and QKD network are used to indicate the communication network based on QKD device. For the sake of better argument, QKD network is defined as a set of infrastructures for generating ITS key based on the laws of quantum mechanics \cite{poppe2008outline} in this paper. The QSCN is defined as a network that provides the secure communication service utilizing the keys generated by QKD network. In order to achieve the ITS secure communication, the  one-time-pad (OTP) encryption algorithm is adopted in the performance analysis in this paper. If ITS is not pursued in an application, the popular encryption algorithms, such as AES, DES and etc., are also acceptable. The QSCN consists of two parts: QKD network and classical network \cite{mehic2017analysis}, shown as Fig. \ref{fig:network}.

Similar to the traditional classical network, QSCN mainly consists of terminals, switches, links, and protocols \cite{mchale1997communication}. The functions of each component are introduced as follows.
\begin{itemize}
\item[-] The terminal is the interface between a user and a communication network, which is mainly used to transmit and receive data. In the QSCN, the terminal is abstracted as a node, which can be a source node for data transmission or a destination node for data reception.

\item[-] The switch is the network device with the function of finding the next receiver and forwarding the data. In the QSCN, the switch needs to be completely trustable and can be abstracted as a trusted relay, which is used to forward data.

\item[-] The link refers to the medium of data communication between terminals. In fact, it includes not only a physical channel but also various communication devices, such as modulators and controllers. In the QSCN, it is abstracted as a P2P channel.

\item[-] The protocol defines the series of rules that enable the network to work properly. In the QSCN, it mainly includes packet protocol and routing protocol.
\end{itemize}
\section{Quantum secure communication network model}
\label{sec:design}
\begin{figure}[htbp]
	\centering
	\includegraphics[scale=0.7]{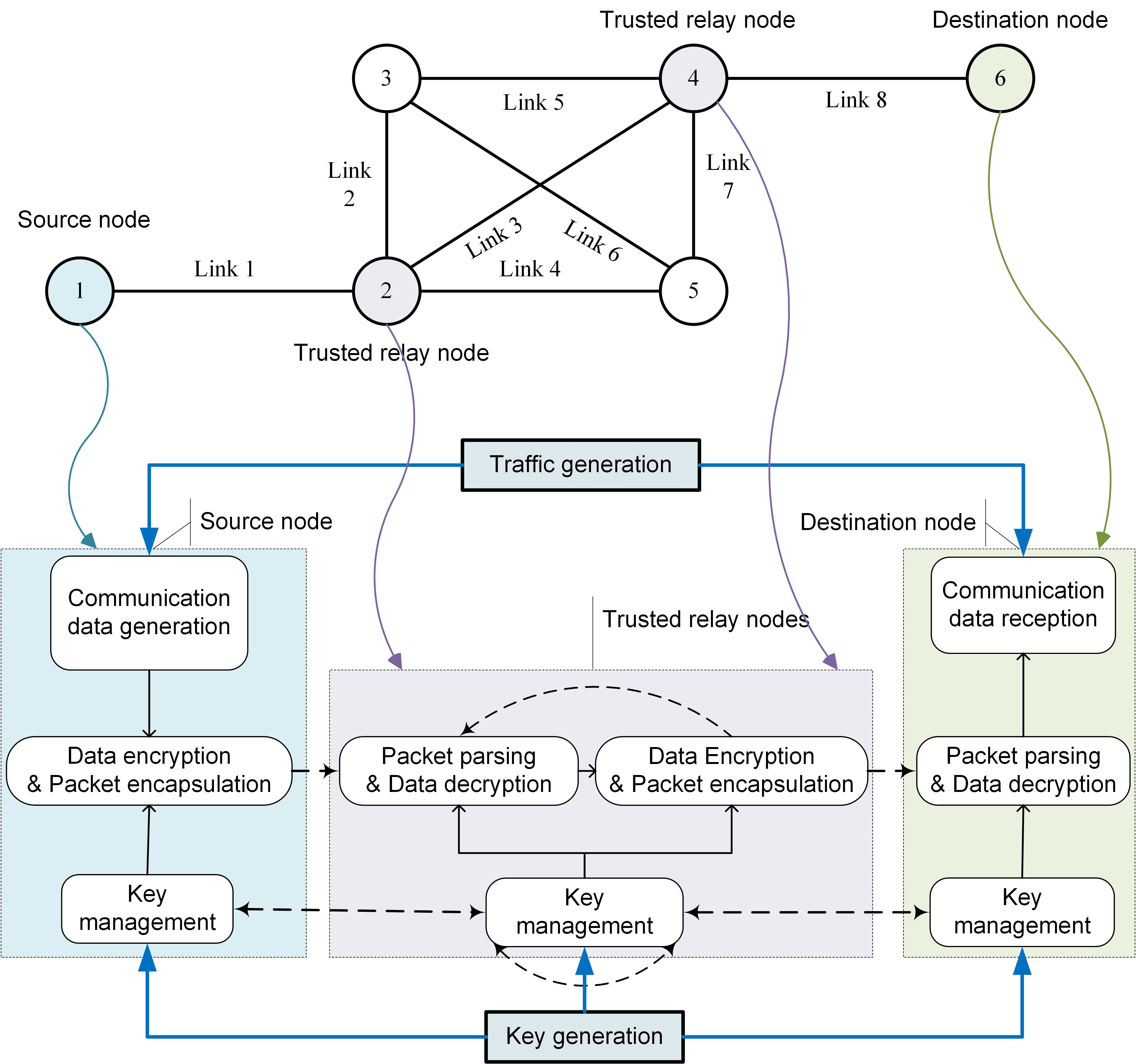}
	\caption{QSCN model}
	\label{fig:QCNA} 
\end{figure}

The biggest difference between the QSCN and classical network is that the classical communication of QSCN consumes the keys generated by QKD layer. This leads to the performance of the QSCN is closely limited by the matching degree between the traffic demand and key generation capability. In fact, the traffic demand in the QSCN is from E2E partners, while the key generation capability is decided by P2P links. In order to describe the network performance more accurately, this paper proposes a practical QSCN model, shown as Fig. \ref{fig:QCNA}.

In the QSCN model, the function of Traffic generation module is to model the E2E traffic demand of classical network, and the function of Key generation module is to model the P2P key generation capability of the QKD network. In addition, the relationship between these two functions will be measured by two proposed performance indicators. The schemes of other modules, such as routing protocols, data encryption/decryption, etc., can be borrowed from the traditional classical network model.

\subsection{Traffic generation module}
\label{subsec:traffic}
Due to the neglect of the relationship between key generation and key consumption, the traffic demand of the classical communication \cite{weigle2006tmix, varet2014generate, bonelli2005brute, ammar2011new, botta2012tool} has not attracted enough attentions in the field of QSCN. A reasonable traffic demand model of the classical network is designed in this section.

The packet transmission process can be assumed to satisfy the following three conditions \cite{weigle2006tmix}:
\begin{enumerate}[(1)]
\item   In the non-overlapping time period, the numbers of transmitted packets are independent variables.

\item   In an arbitrarily short time $\Delta t$, the probability of transmitting a packet is independent of the starting time and only proportional to the length of the time period.

\item   In an arbitrarily short time $\Delta t$, the number of transmitted packets is either 1 or 0.
\end{enumerate}

It can be proved that the packet transmission process follows a Poisson stochastic process \cite{gardiner2009stochastic}. Let $\lambda $ be the average number of transmitted packets per second, and ${p_k}\left( t \right)$ denotes the probability of transmitting $k$ packets within the time period $t$. According to the conditions as above, in an arbitrarily short time $\Delta t$,the average number of transmitted packets is $\lambda \Delta t$. Dividing the finite time period $t$ into $n$ small time slices $\Delta t$, i.e., $t = n\Delta t$, then the $k$ packets transmitted in the time period $t$ can be divided into $n$ parts. Let $n \to \infty $, then the probability of transmitting $k$ packets in time period $t$ is
\begin{equation}
\label{eq:p_k}
{p_k}\left( t \right) = \frac{{{e^{ - \lambda t}}{{\left( {\lambda t} \right)}^k}}}{{k!}},k = 0,1,2,...
\end{equation}

The Eq.\ref{eq:p_k} states that, the probability ${p_k}\left( t \right)$ of transmitting $k$ packets in time period $t$ obeys the Poisson distribution under the above three conditions \cite{gardiner2009stochastic}. According to the basic properties of probability theory, it can be derived that:
\begin{enumerate}[(1)]
\item   The probability density function of the packet transmission interval is $\lambda {e^{ - \lambda t}}$. In other words, the packet transmission interval is subject to the exponential distribution.

\item   The average number of transmitted packets during the time period $t$ is $E\left( {k,t} \right) = \sum\limits_{k = 0}^\infty  {k{p_k}\left( t \right) = \lambda t} $. Therefore, the average number of transmitted packets per second is $E\left( {k,1} \right) = \lambda $, and the average transmitted packets interval is $1/\lambda $. Set the packet size to $\kappa $, the average communication rate is ${R_f} = \lambda \kappa $.

\item   Let $\eta $ denote a random number sequence of multiple packet transmission intervals, which satisfies the exponential distribution with a mean of $1/\lambda $. Therefore, its cumulative distribution function is $F\left( \eta  \right) = 1 - {e^{ - \lambda \eta }}$. Suppose $\xi $ is a random number sequence with uniformly distributed in $\left[ {0,1} \right]$, $\eta $ can be produced as follows,

    \begin{equation}
    \eta  = \frac{1}{\lambda }\ln \left( {1 - \xi } \right)
    \end{equation}

\end{enumerate}

According to the analysis above, if the average number of transmitted packets per second is $\lambda $, the packet transmission process can be simulated into an exponential distribution with the average transmission interval of $1/\lambda $. Therefore, the traffic generation module is constructed by the exponential distribution based packet transmission interval.

\subsection{Key generation module}
\label{subsec:key}
A QKD system consists of the optical sub-system and the post-processing sub-system. The key generation capability is determined by both performance of the two sub-systems \cite{tamaki2018information}. Therefore, it is not convincing to characterize the key generation capability only by the performance of the post-processing procedure as in the literature \cite{mehic2017implementation}.

In 2004, Gottesman et al. proposed the GLLP theory to calculate the secure key generation rate of QKD system with imperfect devices \cite{gottesman2004security}. This method has been adopted in most practical QKD systems \cite{tang2016experimental, zhou2016making, tang2016measurement, lucamarini2018overcoming, rosenberg2007long, zhao2006experimental, gisin2006trojan, sasaki2011field, lo2012measurement, schmitt2007experimental, deng2004secure, braunstein2012side}. Therefore, we use the GLLP theory to model the key generation module. The formula to calculate the secure key generation rate is as follows,
\begin{equation}
\label{R}
R = \max \left\{ { - q{Q_\mu }{f_{ec}}H\left( {{E_\mu }} \right) + q{Q_1}\left[ {1 - H\left( {{e_1}} \right)} \right],0} \right\}.
\end{equation}
In the Eq. \ref{R}, $q$ is the sifting coefficient, the subscript $\mu $ denotes the intensity of signal state, ${Q_\mu }$ is the overall gain of signal state, ${E_\mu }$ is the overall quantum bit error rate (QBER), ${Q_1}$ is the gain of single-photon state, ${e_1}$ is the error rate of single-photon state, ${f_{ec}}$ is the error correction efficiency, and $H\left( x \right)$ is the binary Shannon information function, given by $H\left( x \right) =  - x{\log _2}\left( x \right) - \left( {1 - x} \right){\log _2}\left( {1 - x} \right).$

Four key variables, ${Q_\mu }$, ${E_\mu }$, ${Q_1}$, ${e_1}$, are required to calculate $R$. The first two can be measured directly by experiment, however, the last two can only be estimated by decoy state method. Let ${\eta _{Bob}}$ denote the transmittance of Bob, including the internal transmittance of optical components and detector efficiency. Note that ${Y_0}$ is the background rate and ${e_0}$ is the error rate of the background. Assuming that Alice and Bob choose the vacuum + weak decoy state method, with expected photon numbers $\nu $ and $\phi $, which satisfy $\nu  < \mu $ and $\phi  = 0$. Thus Alice and Bob can estimate the background rate,
\begin{equation}
\label{Q_phi}
{Q_\phi } = {Y_0},
\end{equation}
\begin{equation}
\label{E_phi}
{E_\phi } = {e_0}.
\end{equation}
The overall gain and QBER of the signal state is given by
\begin{equation}
\label{Q_mu}
{Q_\mu } = 1 - {e^{ - \eta \mu }} + {Y_0},
\end{equation}
\begin{equation}
{Q_\mu }{E_\mu } = {e_0}{Y_0} + {e_{det}}\left( {1 - {e^{ - \eta \mu }}} \right).
\end{equation}
The overall gain and QBER of the weak decoy state is given by
\begin{equation}
\label{Q_nu}
{Q_\nu } = 1 - {e^{ - \eta \nu }} + {Y_0},
\end{equation}
\begin{equation}
{Q_\nu }{E_\nu } = {e_0}{Y_0} + {e_{det}}\left( {1 - {e^{ - \eta \nu }}} \right).
\end{equation}

Considering the statistical fluctuations of finite data-set size \cite{ma2005practical, ma2012statistical} in real-life experiments on the estimation process of ${Q_1}$ and ${e_1}$, the Chernoff bound \cite{curty2014finite} is used to provide good bounds \cite{yin2014long}. Let ${X_1}$, ${X_2}$, $...$ , ${X_n}$ be a set of independent Bernoulli random variables that satisfy $P\left( {{X_i} = 1} \right) = {p_i}$, and let $X = \sum\limits_{i = 1}^n {{X_i}} $ and $\overline X  = E\left( X \right) $. With the security bound of $\varsigma = 2\varepsilon $, we can draw the following inequalities \cite{yin2014long},
\begin{equation}
\label{inequalities1}
P\left( {X - \overline X  > \sqrt {2X\ln \left( {{\varepsilon ^{ - 3/2}}} \right)} } \right) \le \varepsilon,
\end{equation}
\begin{equation}
\label{inequalities2}
P\left( {\overline X  - X > \sqrt {2X\ln \left( {16{\varepsilon ^{ - 4}}} \right)} } \right) \le \varepsilon.
\end{equation}
Denote the number of pluses sent by Alice for signal as ${N_\mu }$, for weak as ${N_\nu }$ and for vacuum as ${N_\phi }$. Let $X$ denote the response count of signal state, then $X = {N_\mu }{Q_\mu }$. According to Eq.\ref{inequalities1} and Eq.\ref{inequalities2}, we can draw that
\[P\left( {{N_\mu }{Q_\mu } - \sqrt {2{N_\mu }{Q_\mu }\ln \left( {{\varepsilon ^{ - 3/2}}} \right)}  \le {N_\mu }{{\overline Q}_\mu } \le {N_\mu }{Q_\mu } + \sqrt {2{N_\mu }{Q_\mu }\ln \left( {16{\varepsilon ^{ - 4}}} \right)} } \right) \ge 1 - \varsigma \]
Therefore, the lower bound and upper bound of ${\overline Q_\mu }$ is given by
\begin{equation}
\overline Q_\mu ^L = {Q_\mu } - \sqrt {\frac{{2{Q_\mu }\ln \left( {{\varepsilon ^{ - 3/2}}} \right)}}{{{N_\mu }}}},
\end{equation}
\begin{equation}
\overline Q_\mu ^U = {Q_\mu } + \sqrt {\frac{{2{Q_\mu }\ln \left( {16{\varepsilon ^{ - 4}}} \right)}}{{{N_\mu }}}}.
\end{equation}
And so on, for the $\overline Q_\nu  ^L$, $\overline Q_\nu ^U$, $\overline Q_\phi ^L$ and $\overline Q_\phi ^U$.

Let $X$ denote the error count of weak decoy state, then $X = {N_\nu }{Q_\nu }{E_\nu }$. According to Eq.\ref{inequalities1} and Eq.\ref{inequalities2}, the lower bound and upper bound of $\overline E_\mu $ is given by
\begin{equation}
\overline E_\nu ^L = {E_\nu } - \sqrt {\frac{{2{E_\nu }\ln \left( {{\varepsilon ^{ - 3/2}}} \right)}}{{{N_\nu }{Q_\nu }}}},
\end{equation}
\begin{equation}
\overline E_\nu ^U = {E_\nu } + \sqrt {\frac{{2{E_\nu }\ln \left( {16{\varepsilon ^{ - 4}}} \right)}}{{{N_\nu }{Q_\nu }}}}.
\end{equation}
The $\overline E_\phi ^L$ and $\overline E_\phi ^U$ can be deduced by analogy.

Combining Eq.\ref{Q_mu} and Eq.\ref{Q_nu}, the lower bound of ${Y_1}$ is given by
\begin{equation}
{Y_1} \ge Y_1^L = \frac{\mu }{{\mu \nu  - {\nu ^{\rm{2}}}}}\left( {{\overline Q_\nu ^L}{e^\nu } - \frac{{{\nu ^2}}}{{{\mu ^2}}}{\overline Q_\mu ^U}{e^\mu } - \frac{{{\mu ^{\rm{2}}}{\rm{ - }}{\nu ^{\rm{2}}}}}{{{\mu ^{\rm{2}}}}}{Y_0}} \right).
\end{equation}
Under the condition of  Eq.\ref{Q_phi} and Eq.\ref{E_phi}, the lower bound of ${Q_1}$ and upper bound of ${e_1}$ is given by
\begin{equation}
{Q_1} \ge Q_{_1}^L = \frac{{{\mu ^2}{e^{ - \mu }}}}{{\mu \nu  - {\nu ^{\rm{2}}}}}\left( {{\overline Q_\nu ^L}{e^\nu } - \frac{{{\nu ^2}}}{{{\mu ^2}}}{\overline Q_\mu ^U}{e^\mu } - \frac{{{\mu ^{\rm{2}}}{\rm{ - }}{\nu ^{\rm{2}}}}}{{{\mu ^{\rm{2}}}}}{\overline Q_\phi ^U}} \right),
\end{equation}
\begin{equation}
{e_1} \le e_1^U = \frac{{\mu \nu  - {\nu ^2}}}{{\mu \nu }}\frac{{{\overline Q_\nu ^U}{\overline E_\nu ^U}{e^\nu } - {\overline Q_\phi ^L}{\overline E_\phi^L}}}{{{\overline Q_\nu ^L}{e^\nu } - \frac{{{\nu ^2}}}{{{\mu ^2}}}{\overline Q_\mu ^U}{e^\mu } - \frac{{{\mu ^2} - {\nu ^2}}}{{{\mu ^2}}}{\overline Q_\phi ^U}}}.
\end{equation}
Substituting these parameter estimations into Eq.\ref{R}, we can calculate the lower bound of the secure key generation rate of one QKD device
\begin{equation}
R \ge \max \left\{ {{R_L} =  - q{Q_\mu }{f_{ec}}H\left( {{E_\mu }} \right) + qQ_{_1}^L\left[ {1 - H\left( {e_{_1}^U} \right)} \right],0} \right\}.
\end{equation}
Then, the key generation capability of a P2P link is given by
\begin{equation}
{R_k} = \max \left\{ {{f_{req}}{R_L},0} \right\},
\end{equation}
where ${f_{req}}$ is the repetition rate.

\subsection{Performance indicators}
\label{subsec:indicator}
To meet the requirement of ITS, it is necessary to utilize the OTP algorithm in the QSCN for data encryption and decryption. The key generation capability of the network is based on the P2P links, however the traffic demand is based on the E2E partners. Under this condition, if the key generation capability cannot meet the traffic demand, i.e. if the QSCN will paralyze at some point, directly affects the performance of a QSCN. However, there is no counterpart in classical network, and there is no appropriate indicator in classical network to measure such performance of QSCN either. In this paper, two performance indicators are proposed: ITS communication capability and ITS communication efficiency, as a necessary supplements to evaluate a QSCN except for those traditional performance indicators of classical networks.
\subsubsection{ITS communication capability}
For a given topology $G = \left( {V,E} \right)$, let $R_k^m\left( {m \in E} \right)$ be the key generation capability of the link $m$ which is time-independent and can be calculated according to GLLP theory. ${R_f}$ indicates the total traffic demand of all pairs of partners and ${R_f} \cdot {\overline P^{i,j}}\left( {i \in V,j \in V} \right)$ means the average traffic demand of node $i$ and node $j$, which can be modeled by a Poisson stochastic process and can be calculated according to the average packet interval. ${\overline \omega ^{i,j,m}} \in \{ 0,1\} $ indicates whether the communication of node $i$ and node $j$ requires link $m$. In addition, ${\overline O^m}$ represents the average number of keys consumed by routing data on the link $m$.

To guarantee the stable operation of the network, the key consumption of each link needs to be less than or equal to its key generation. The key consumption mainly includes communication data consumption and routing data consumption. When the OTP algorithm is used, the key consumption of communication data is equal to the traffic.

\begin{equation}
\forall m \in E,{\text{ }}R_k^m \geqslant \sum\limits_{i \in V} {\sum\limits_{j \in V} {{R_f} \cdot {{\overline P}^{i,j}}{{\overline \omega }^{i,j,m}} + {{\overline O}^m}} }
\end{equation}

The ITS communication capability of the QSCN is defined as the maximum traffic demand that enables all links to work stably. It is represented by a symbol $C$.

\begin{equation}
\label{eq:capability}
C = \max \left\{ {{R_f}{\text{ }}|{\text{ }}\forall m \in E,{\text{ }}\sum\limits_{i \in V} {\sum\limits_{j \in V} {{R_f} \cdot {{\overline P}^{i,j}}{{\overline \omega }^{i,j,m}} + {{\overline O}^m}} }  - R_k^m \leqslant 0} \right\}
\end{equation}

It can be seen from the formula that the ITS communication capability is mainly related to the traffic demand, key generation capability and routing protocol.

\subsubsection{ITS communication efficiency}
When the traffic demand is higher than the ITS communication capability, the QSCN will inevitably paralyze after running for a certain period of time. The time span before the QSCN paralyzes is defined as ITS operation time, represented by the symbol ${T_o}$. Let $D$ be the initial number of keys on each link in the network and ${R_f} \cdot {P^{i,j}}\left( t \right)\left( {i \in V,j \in V} \right)$ be the traffic demand of node $i$ and node $j$ at the time of $t$. In addition, ${\omega ^{i,j,m}}\left( t \right) \in \{ 0,1\} $ indicates whether the communication of node $i$ and node $j$ requires link $m$ at the time of $t$ and ${O^m}\left( t \right)$ indicates the number of keys consumed by routing data on the link $m$ at the time of $t$. Then the network ITS operation time must meet the following requirement:

\begin{small}
\begin{equation}
\label{eq:operation}
{T_o} = \max \left\{ {T{\text{ }}|{\text{ }}\forall m \in E,{\text{ }} \int\limits_0^T {\left( {\sum\limits_{i \in V} {\sum\limits_{j \in V} {{R_f} \cdot {P^{i,j}}\left( t \right){\omega ^{i,j,m}}\left( t \right) + {O^m}\left( t \right)} } } \right)} dt
- T \cdot R_k^m \leqslant D} \right\}
\end{equation}
\end{small}

	When the network paralyzes, the number of remaining keys on the link $m$ is:

\begin{equation}
{D_m} = \int\limits_0^{{T_o}} {\left( {\sum\limits_{i \in V} {\sum\limits_{j \in V} {{R_f} \cdot {P^{i,j}}\left( t \right){\omega ^{i,j,m}}\left( t \right) + {O^m}\left( t \right)} } } \right)} dt - {T_o} \cdot R_k^m
\end{equation}

The ITS recovery time is defined as the length of time required for the numbers of keys on all links recover to $D$, represented by the symbol ${T_r}$.

\begin{equation}
\label{eq:recovery}
{T_r} = \min \left\{ {T{\text{ }}|{\text{ }}\forall m \in E,T \geqslant \frac{{D - {D_m}}}{{R_k^m}}} \right\}
\end{equation}

In order to ensure stable operation based on the OTP algorithm in any traffic demand environment, the ITS operation time and ITS recovery time of the QSCN must meet the preset requirement. The ratio of the ITS operation time to the sum of ITS operation time and ITS recovery time is defined as ITS communication efficiency $Q$ as in the Eq. \ref{eq:efficiency}.

\begin{equation}
\label{eq:efficiency}
Q = \frac{{{T_o}}}{{{T_o} + {T_r}}}
\end{equation}

The ITS communication efficiency, which is mainly related to the traffic demand, key generation capability and routing protocol, can be used to guide the actual working policy of the QSCN.

It can be seen from the formulas that when the network scale becomes large, it is quite difficult to theoretically analyze the QSCN performance. In this case, the network simulation becomes the most important, convenient and economical solution for the performance analysis of the QSCN.

\section{Simulation and results analysis}
\label{sec:simulation}
In this section, a QSCN simulation using Network Simulator-version 3 (NS-3) \cite{henderson2008network} is designed to analyze the network performance.
\subsection{Simulation design}
\label{subsec:design}
The simulation design mainly includes the traffic generation, key generation, topology and routing protocol, which can directly affect the performance of a QSCN.
\subsubsection{Traffic generation}
To simplify the analysis, the traffic demand between any two partners in our simulation are assumed to be the same scale. Based on the analysis in Sec. \ref{subsec:traffic}, the packet transmission intervals follow an exponential distribution. In order to find out the performance bottleneck of QSCN , two comparison simulations are conducted with communication rates of 100 Kbps and 10 Kbps. In the case where the packet size is set to 500 bytes, the average packet transmission intervals are set to 40 milliseconds and 400 milliseconds.

In a practical network, a classical link often needs to serve multiple partners, which will lead to two kinds of problems. Firstly, when the traffic demand of a pair of terminal partners is met, it may reduce the communication performance of another pair of terminal parties. Secondly, the change of paths for a certain pair of terminal partners may improve the communication performance of others. Therefore, it is not reasonable to only analyze the traffic demand of one pair of terminal partners as in the literature \cite{mehic2017implementation}. Therefore, we should consider the traffic demand of all pairs of partners at the same time, as shown in Algorithm \ref{algorithm:traffic}.

\renewcommand{\algorithmicrequire}{\textbf{Input:}} 
\renewcommand{\algorithmicensure}{\textbf{Output:}} 

\begin{algorithm}[h]
	\caption{End-to-end traffic generation of whole network}
    \label{algorithm:traffic}
	\begin{algorithmic}[1] 
		\Require a given topology $G = \left( {V,E} \right)$ with the node size of $n$ and the edge size of $m$, traffic generation module with the average packet interval of $1/\lambda $, packet size of $\kappa $, routing protocol of $\omega$, total duration of simulation $T$
		\Ensure The key consumptions of all links  $comCnt$ (an array of size $m$)
		\State $comCnt \gets 0$
        \While {$ t < T $}
		\For {each vertex $ v_i, v_j(1 \leqslant i \leqslant n,1 \leqslant j \leqslant n , i \neq j) \in G,V $}
		\For {each edge $ e_k(1 \leqslant k \leqslant m) \in G,E $}
        \State $comCnt\left( k \right) \gets comCnt\left( k \right) + {\omega ^{i,j,m}}\left( t \right) \cdot \kappa $ \Comment{Calculate the key consumption of the link $e_k$}
        \State $\xi  = Rnd\left( {0,1} \right)$ \Comment{Let $\xi$ as a random number in the $[0,1]$}
        \State $ \eta  = \frac{1}{\lambda }\ln \left( {1 - \xi } \right){\text{ }} $ \Comment{Calculate the packet interval}
		\State $ t \gets t + \eta $  \Comment{Send next packet after the packet interval}
		\EndFor
		\EndFor
		\EndWhile
	\end{algorithmic}
\end{algorithm}

\subsubsection{Key generation}
To simplify the analysis, we assume that only one QKD device is configured on each link and the parameters of all QKD devices are the same. The parameters of QKD device in our simulation are shown in Table \ref{tab:key}, referring to the literature \cite{yuan201810}. It should be noted that the QSCN model supports the configuration of multiple QKD devices for each link and the different parameter setting for each QKD device.

\begin{table}[htbp]
	\centering
	\caption{Parameters of QKD devices}
	\label{tab:key}       
	\begin{threeparttable}
    \renewcommand\arraystretch{1.5}
    \setlength{\tabcolsep}{1.2mm}{
        \begin{tabular}{ccccccccccccccc}
    		\toprule[1pt]
    		${f_{req}}$ & $q$ & $\alpha $ & $\eta_{Bob}$ & $e_{det}$ & $\mu$ & $\nu$ & $\phi$ & $Y_0$ & $e_0$ & ${f_{ec}}$ & ${N_{\mu}}$ & ${N_{\nu}}$ & ${N_{\phi}}$ & $\varsigma$ \\
    		\midrule
    		1 GHz & 0.9 & 0.2 db/km & 0.1 & 0.01 & 0.4 & 0.1 & 0 & 2.1E-5 & 0.5 & 1.15 & 1.6E10 & 2E9 & 2E9 & 5.73E-7\\
    		\bottomrule[1pt]
    	\end{tabular}
    }
    \end{threeparttable}
\end{table}

According to the parameters as Table \ref{tab:key}, the key generation rate of each link can be calculated. In order to find out the performance bottleneck of the whole network, we make all QKD devices with different distribution distances start generating keys at the same time, and simulate the process of key generation, as shown in Algorithm \ref{algorithm:key}.

\begin{algorithm}[h]
	\caption{Point-to-point key generation of whole network}
    \label{algorithm:key}
	\begin{algorithmic}[1] 
		\Require a given topology $G = \left( {V,E} \right)$ with the node size of $n$ and the edge size of $m$, key generation module with $\left( {{f_{rep}},q,{Q_\mu },{E_\mu },{f_c},{Q_1},{e_1}} \right)$, total duration of simulation $T$
		\Ensure The key generations of all links $genCnt$ (an array of size $m$)
		\State $genCnt \gets 0$
		\For {each link $ e_k(1 \leqslant k \leqslant m) \in G,E $}
		\State ${R_k} \gets \max \left\{ { - {f_{req}}q{Q_\mu }{f_{ec}}H\left( {{E_\mu }} \right) + {f_{req}}qQ_{_1}^L\left[ {1 - H\left( {e_{_1}^U} \right)} \right],0} \right\}$ \Comment{Calculate the key generation rate}
		\State $ genCnt\left( k \right) \gets genCnt\left( k \right) + {R_k} \cdot T $
		\EndFor
	\end{algorithmic}
\end{algorithm}

\subsubsection{Topology}
\begin{figure}[htbp]
	\centering
	\includegraphics[scale=0.7]{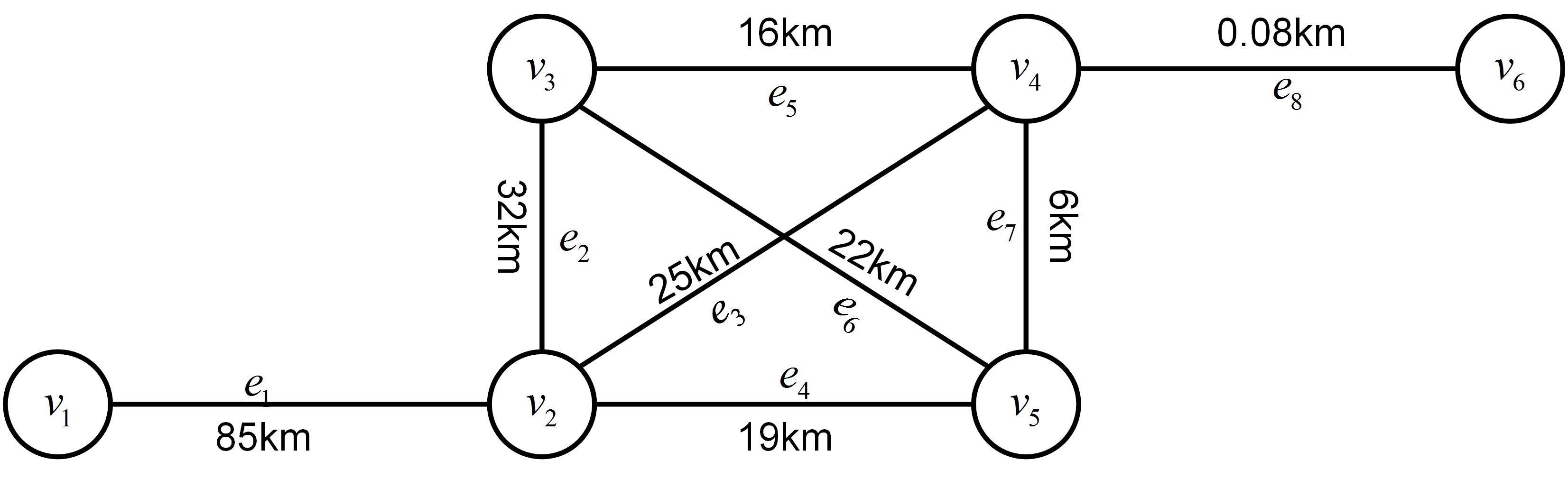}
	\caption{Topology of the QSCN}
	\label{fig:topology}       
\end{figure}

To the best of our knowledge, the most comprehensive study on QSCN performance appears in the literature \cite{mehic2017implementation}. For comparison, the topology of SECOQC network \cite{poppe2008outline} used in \cite{mehic2017implementation}, as shown in Fig. \ref{fig:topology}. Besides, the trusted relays adopted in our simulation work on the basis of the hop-by-hop \cite{poppe2008outline}. Each node in the topology is a communication user and acts as a trusted relay. In another word, each node should be configured with the classical communication equipment, the QKD device, encryption/decryption module, key management module, traffic monitoring module.

\subsubsection{Routing protocol}
The traffic demand that the network needs to meet is E2E based, however, the key consumption is P2P based. Through the design of better routing protocol, it is expected to maximize the utilization of the P2P key generation capability to satisfy the E2E traffic demand, thereby improving network performance. Due to the frequent changes of the number of keys during the simulation process, the connected/broken state of a link may change as well, which leads to frequent changes of the network topology. It means that the selected routing protocol should be able to perceive the network topology changes in time. Therefore, the DSDV routing protocol \cite{perkins1994highly}, which is commonly used in the wireless ad hoc network, is adopted. The DSDV protocol updates routing table regularly.

\subsection{Performance of traditional indicators}
\label{subsec:traditional}
In this section, four most important performance indicators of classical network are evaluate to analyze the QSCN performance, which are: one-way delay (OWD) \cite{almes2016one}, throughput \cite{burgess2004rfc}, packet delivery rate (PDR) \cite{mehic2017implementation} and routing cost (RCost) \cite{evans2009routing}. OWD is the required time of data packet transmission across the network, which may be affected by any component of the related links. Throughput is defined as the rate of successful message delivery over a communication channel. PDR refers to the ratio of the received packets by the destination to the generated packets by the source. RCost refers to the amount of routing data generated during network operation.

\begin{figure}[htbp]
	\centering
	\includegraphics[scale=0.075]{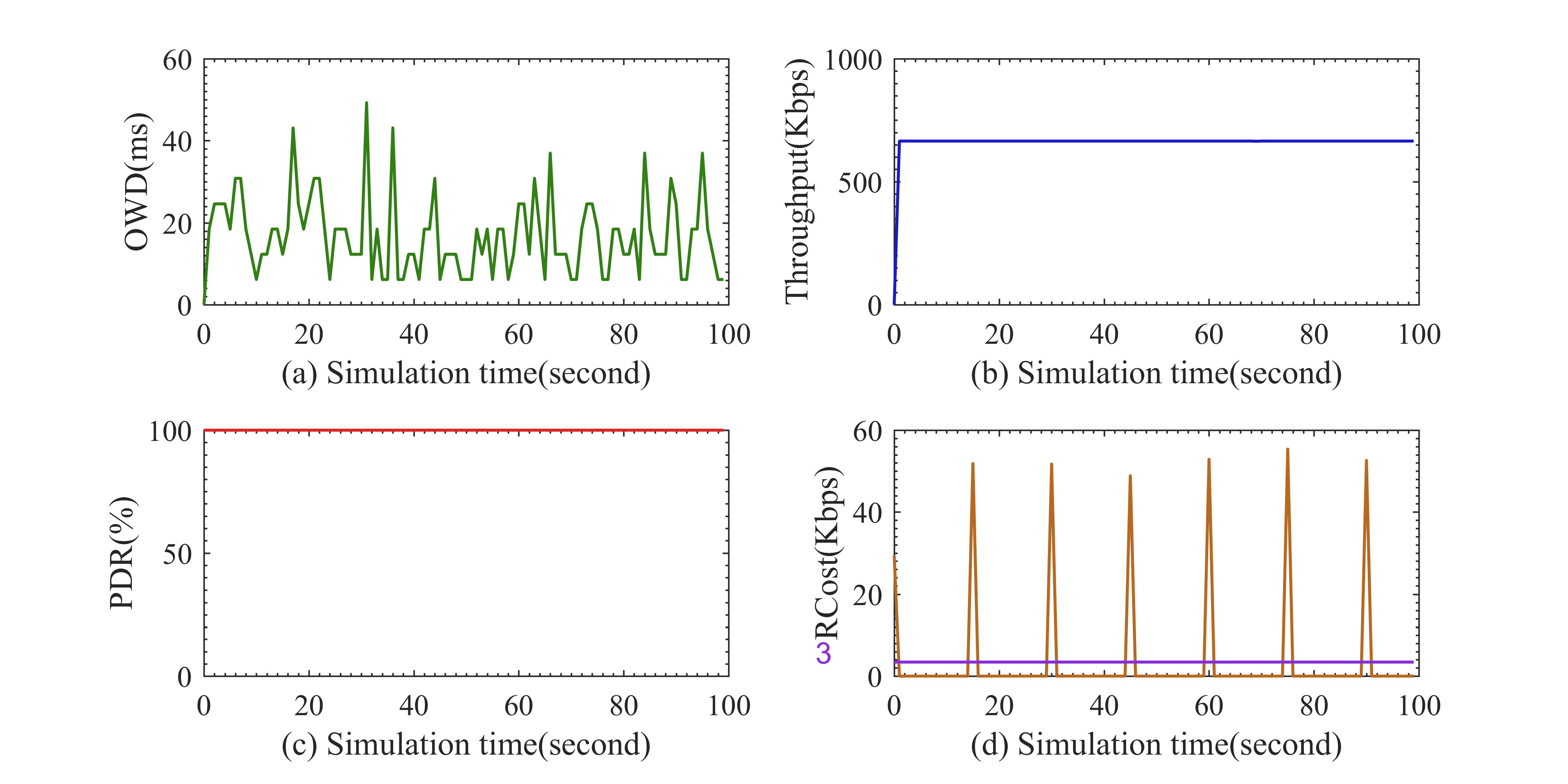}
	\caption{Network performance with 10 Kbps communication rate}
	\label{fig:10Kbps}       
\end{figure}

\begin{figure}[htbp]
	\centering
	\includegraphics[scale=0.075]{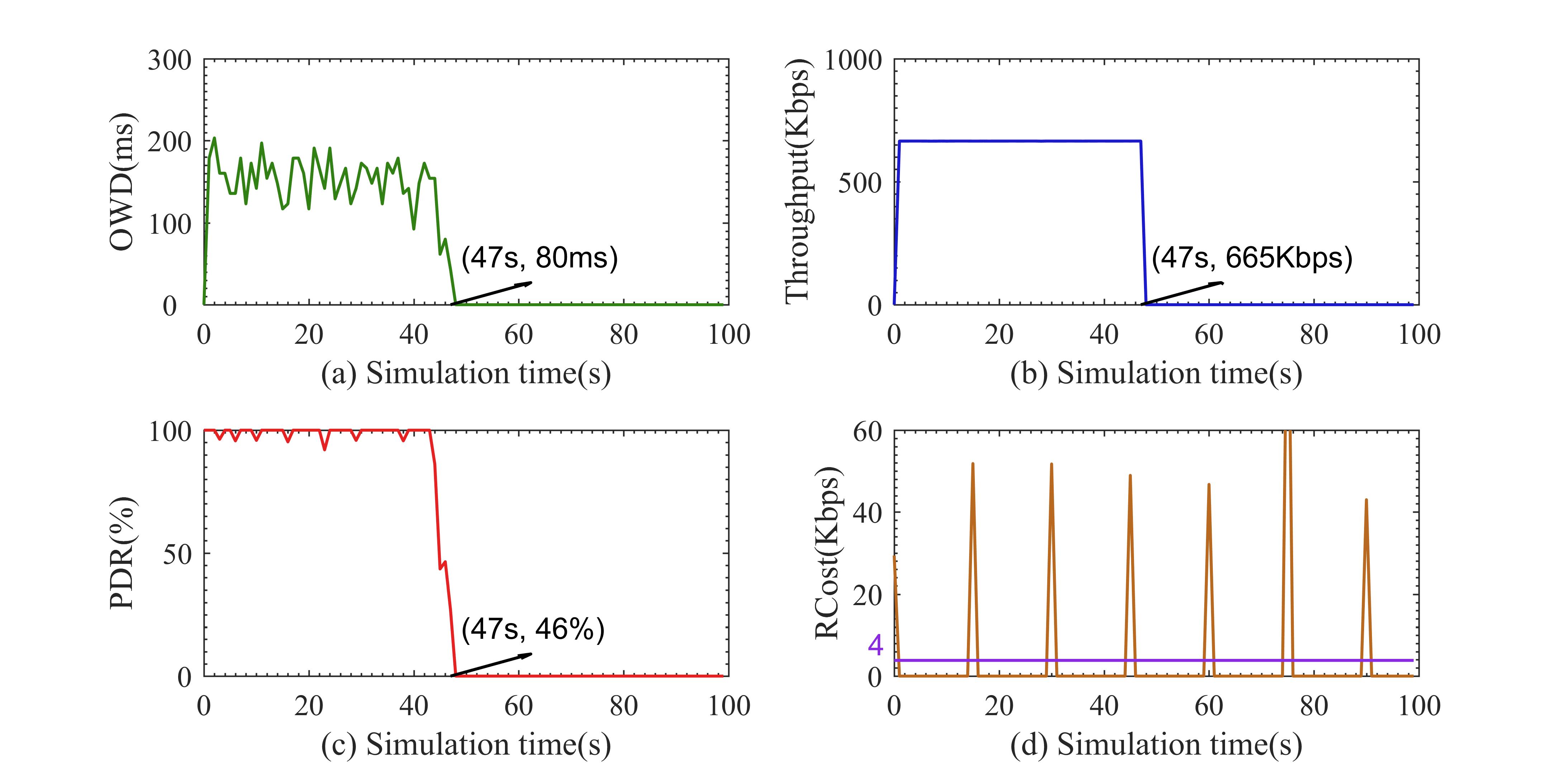}
	\caption{Network performance with 100 Kbps communication rate}
	\label{fig:100Kbps}       
\end{figure}

Based on the QSCN model and the foregoing parameters, the simulation results of OWD $\left( {{v_1} \to {v_6}} \right)$, throughput $\left( {{v_1} \to {v_6}} \right)$, PDR $\left( {{v_1} \to {v_6}} \right)$, and RCost of whole network with the average packet interval of 400 millisecond (communication rates of 10 Kbps) and 40 millisecond (100 Kbps) are shown in Fig. \ref{fig:10Kbps} and Fig. \ref{fig:100Kbps} respectively.

From Fig. \ref{fig:10Kbps} and Fig. \ref{fig:100Kbps}, it can be seen that there is almost no difference in OWD, throughput, and PDR between the two communication rates before the 47th second. However, after the 47th second, the performance in Fig. \ref{fig:10Kbps} remains stable. Conversely, the sharp rise of OWD, sharp drop of throughput and PDR in Fig. \ref{fig:100Kbps} indicates that the node $v_6$ cannot accept any packet from the node $v_1$. In other words, the network is paralyzed at the 47th second.

\subsection{Performance of proposed indicators}
\label{subsec:new}
\subsubsection{ITS communication capability}
\label{subsec:capability}
The simulation results of traditional evaluation indicators show that when the traffic demand exceeds a certain value, the network will paralyze after a certain period of continuous operation. To find out the maximum traffic demand that the QSCN can support stably is very important for the QSCN designer and manager. Therefore, ITS communication capability is proposed and simulated.

\begin{figure}[htbp]
	\centering
	\subfigure[100 Kbps communication rate]{
	\label{fig:100K_compare}
	\includegraphics[scale=0.075]{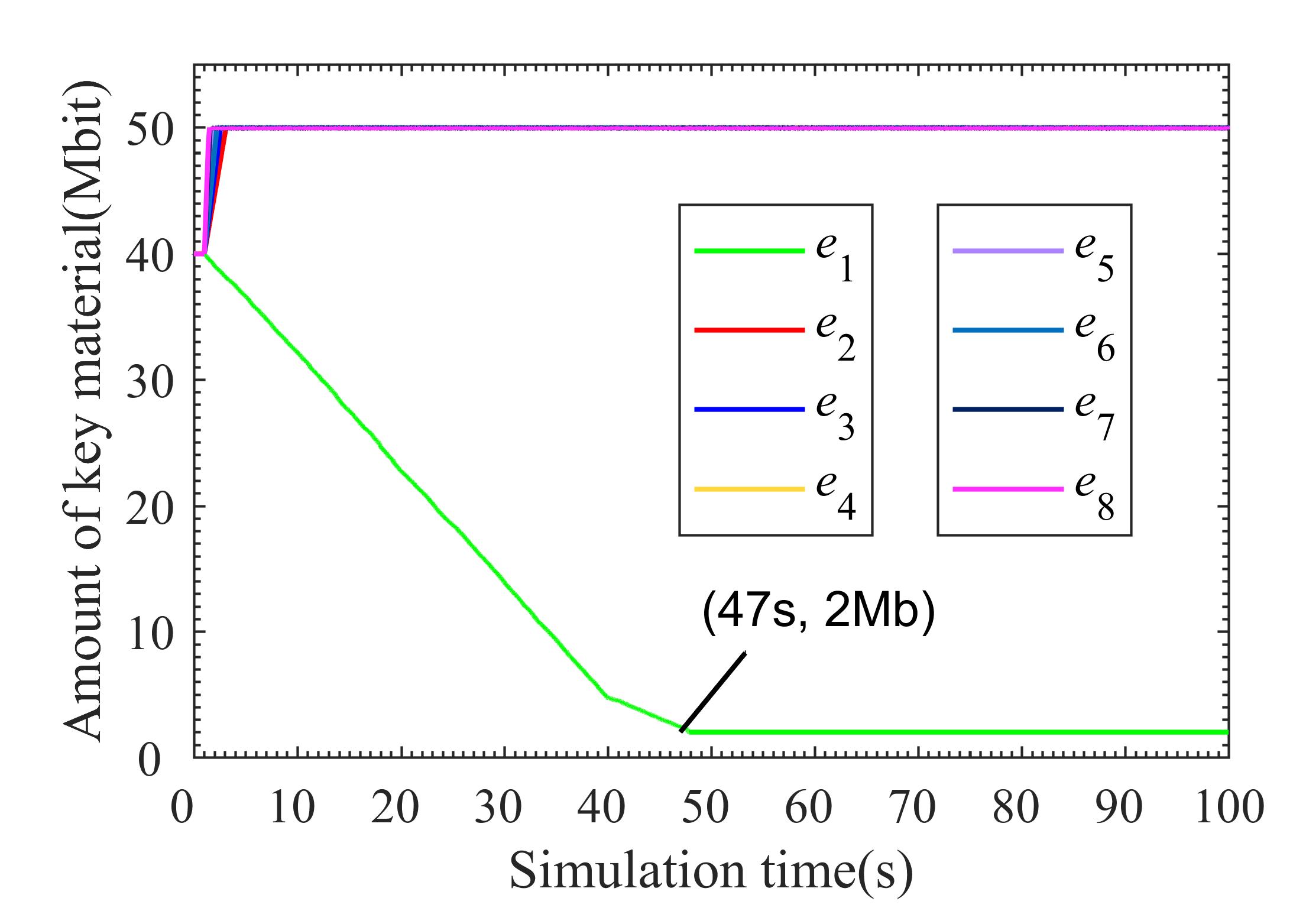}}
	\subfigure[23 Kbps communication rate]{
	\label{fig:23K_compare}
	\includegraphics[scale=0.075]{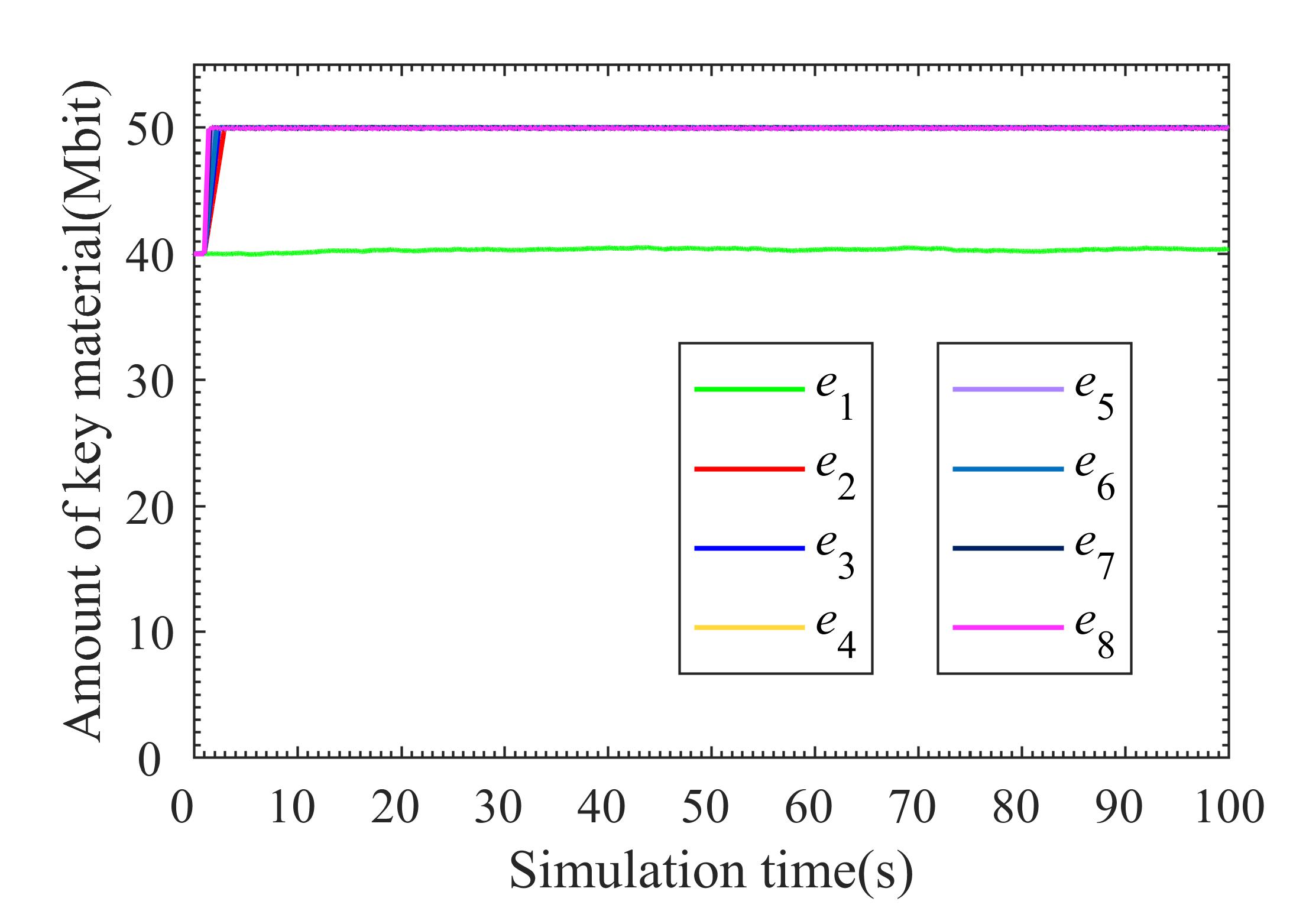}}
	\caption{Key consumption in the SECOQC topology}
	\label{fig:original}
\end{figure}

In order to explore the reasons for paralysis, the key consumption at the communication rate of 100 Kbps is simulated, as shown in Fig. \ref{fig:100K_compare}. Each curve in the figure reflects the key consumption process of a link. Meanwhile, the slope of the curve represents the consumption rate. The initial number of keys of each key pool is set to be 40 Mb. The partners need a small number of pre-shared keys before a new key generation process is established \cite{dodson2009updating, cederlof2008security}. The pre-shared keys are used to guarantee the integrity of the protocol in the first transaction and it should not be used for any other purposes except to establish a new key generation process. In our simulation, the minimal threshold is set to 2 Mb. From the Fig. \ref{fig:100K_compare}, it can be concluded that the link with the fastest key consumption rate is ${e_1}$. At the 47th second, because the remaining number of keys is below the minimum threshold, this link is ``broken''.

DSDV routing protocol finds the optimal path based on the principle of minimum hop count. Therefore, the key generation of link ${e_1}$ needs to satisfy the communication traffic demand of five pairs of communication partners $\left( {{v_1},{v_2}} \right)$, $\left( {{v_1},{v_3}} \right)$, $\left( {{v_1},{v_4}} \right)$, $\left( {{v_1},{v_5}} \right)$ and $\left( {{v_1},{v_6}} \right)$. Considering that the communication process is bidirectional, the link $e_1$ will load 5 * 2 = 10 times of the one-way traffic demand of one pair of communication partners. According to the topology in Fig. \ref{fig:topology}, the length of link $e_1$ is 85 km. By substituting the length into the GLLP theory, the calculated key generation rate is about 233 Kbps. From Fig. \ref{fig:100Kbps}, the consumption rate of DSDV protocol routing data of whole network is about 4 Kbps. There are 8 links in the network, we can deduce that the routing data consumption of link $e_1$ is about ${\text{4/8}} \approx {\text{0.5}}$ Kbps. Therefore, in order to implement ITS transmission in the QSCN directly, the traffic demand that link $e_1$ can afford is only about ${\text{(233 - 0.5)/(5*2)}} \approx {\text{23}}$ Kbps.

Fig. \ref{fig:23K_compare} shows the key consumption when the communication rate is 23 Kbps. It can be seen that the number of keys of link $e_1$ maintains the original amount basically. The results demonstrate that the key consumption and key generation on the link $e_1$ are balanced, which is consistent with the theoretical analysis as above.

\subsubsection{ITS Communication Efficiency}
\label{subsec:paralysis}
ITS communication capability can indicate the maximal traffic demand that the QSCN is able to support stably. However, when the traffic demand exceeds the ITS communication capability, the QSCN must need a certain recovery time after it operate for some time to maintain a stable communication. The less proportion of recovery time means higher performance. Therefore, the ITS communication efficiency is proposed and simulated, which is the ratio of the ITS operation time to the sum of ITS operation time and ITS recovery time.

The simulation results previously show that, due to the insufficient key generation capability of the QSCN, the communication process can only last 47 seconds at the communication rate of 100 Kbps. It needs to make clear that the 47 seconds is the network ITS operation time at the communication rate of 100 Kbps.

\begin{equation}
\label{eq:T_p}
{T_o} \approx 47{\text{ }}seconds
\end{equation}

The simulation results previously show that the first paralyzed link is the link $e_1$. It can be seen from the topology that the key generation rate of the link $e_1$. Therefore, in this simulation, the ITS recovery time of the network depends on the ITS recovery time of the link $e_1$, which can be calculated as Eq.\ref{eq:T_r}.

\begin{equation}
\label{eq:T_r}
{T_r} = \frac{{40{\text{ }}Mb - 2Mb}}{{233{\text{ }}Kbps}} \approx 163{\text{ }}seconds
\end{equation}

According to the Eq.\ref{eq:efficiency}, the ITS communication efficiency can be calculated as Eq.\ref{eq:Q}.

\begin{equation}
\label{eq:Q}
Q = \frac{{{T_o}}}{{{T_o} + {T_r}}} = \frac{{47{\text{ }}seconds}}{{{\text{47 }}seconds + 163{\text{ }}seconds}} \approx 22\%
\end{equation}

From the analysis above, the ITS communication capability and ITS communication efficiency of this QSCN is 23 Kbps and 22\% respectively. Therefore, the QSCN can work stable when the traffic demand is lower than 23 Kbps, such as short message, audio communication, etc. In addition, when the traffic demand is higher than the ITS communication capability, such as real-time video communication, suitable key management mechanism or QKD network improvement need to be designed accordingly.

\section{Conclusion}
\label{sec:conclusion}
In this paper, a practical QSCN model is proposed and the three major improvements include: (I) the volatility of traffic demand of classical network is modeled by the Poisson stochastic process; (II) the capability of key generation in QKD network is calculated by the GLLP theory; (III) two performance indicators are proposed, which are ITS communication capability and ITS communication efficiency. In addition, the simulation is designed based on the QSCN model with the topology of the SECOQC network and the DSDV routing protocol. The plentiful simulation results verified the accuracy of the proposed QSCN model, and the necessity of the proposed performance indicators. In the further, we plain to design better QKD device deployment and routing protocols to improve the ITS communication capability and the ITS communication efficiency.

\begin{acknowledgements}
This work is supported by the National Natural Science Foundation of China
(Grant Number: 61771168, 61702224), Space Science and Technology Advance Research Joint Funds (6141B061\\10105). \end{acknowledgements}

\bibliographystyle{spmpsci} 
\bibliography{reference}

\end{document}